\documentclass[aps,prl,twocolumn,superscriptaddress]{revtex4-1}
\usepackage{graphicx}
\usepackage{gensymb}
\usepackage[intlimits]{amsmath}
\usepackage{amssymb}
\usepackage{soul}
\usepackage{soul}
\begin{document}

\title{Polarization Controlled Directional Scattering for Nanoscopic Position Sensing}
\author{M. Neugebauer}
\author{P. Wo{\'z}niak} 
\author{A. Bag}
\affiliation{Max Planck Institute for the Science of Light, G{\"u}nther-Scharowsky-Str.1, D-91058 Erlangen, Germany}
\affiliation{Institute of Optics, Information and Photonics, Department of Physics, Friedrich-Alexander-University Erlangen-Nuremberg, Staudtstr. 7/B2, D-91058 Erlangen, Germany}
\author{G. Leuchs}
\author{P. Banzer}
\email{peter.banzer@mpl.mpg.de}
\affiliation{Max Planck Institute for the Science of Light, G{\"u}nther-Scharowsky-Str.1, D-91058 Erlangen, Germany}
\affiliation{Institute of Optics, Information and Photonics, Department of Physics, Friedrich-Alexander-University Erlangen-Nuremberg, Staudtstr. 7/B2, D-91058 Erlangen, Germany}
\affiliation{Department of Physics, University of Ottawa, 25 Templeton, Ottawa, Ontario, K1N 6N5 Canada}
\date{\today}
\begin{abstract}
Controlling the propagation and coupling of light to sub-wavelength antennas is a crucial prerequisite for many nanoscale optical devices. Recently, the main focus of attention has been directed towards high-refractive-index materials such as silicon as an integral part of the antenna design. This development is motivated by the rich spectral properties of individual high-refractive-index nanoparticles. Here, we take advantage of the interference of their magnetic and electric resonances, to achieve remarkably strong lateral directionality. For controlled excitation of a spherical silicon nanoantenna we use tightly focused radially polarized light. The resultant directional emission depends on the antenna's position relative to the focus. This approach finds application as a novel position sensing technique, which might be implemented in modern nanometrology and super-resolution microscopy setups. We demonstrate in a proof-of-concept experiment, that a lateral resolution in the \AA{}ngstrom regime can be achieved.
\end{abstract}
\maketitle
\section{Introduction}
Cylindrical vector beams are well established tools in modern microscopy, ranging from scanning microscopy, where a reduced focal spot size can be achieved with a radially polarized beam \cite{Quabis2000,Youngworth2000,Wang2008,Chen2013}, to more sophisticated techniques such as stimulated-emission-depletion \cite{Rittweger2009,Hao2010} and multi-photon microscopy \cite{Yoshiki2005}. In addition, those polarization tailored beams have also paved the way towards versatile applications in recent nanophotonic experiments by enabling selective excitation of nanoparticle eigen-modes \cite{Sancho-Parramon2012,Wozniak2015}, or controllable directional emission and waveguide-coupling of single plasmonic nanoantennas \cite{Neugebauer2014}.\\
In this work, we combine several aspects of both research fields to present a novel approach towards high-precision position sensing, a discipline, which is of paramount importance in modern nanometrology \cite{Gelles1988,Nugent-Glandorf2004,Rohrbach2002,Gittes1998,Dupont2011,Weisenburger2014,Roy2015,Bon2015}, because of its special role in super-resolution microscopy \cite{Hell1994,Klar1999,HessGirirajanTMasonM2006,Betzig2006}. Our all-optical technique for localization of a single nanoantenna is thereby based on encoding the position of the antenna in its laterally directional scattering pattern. For that purpose, we take advantage of the resonance properties of a high-refractive-index silicon nanoantenna featuring electric and magnetic resonances \cite{Evlyukhin2012,Shi2012,Bakker2015,Albella2015}.
\section{Results}
\begin{figure}[htbp]
\includegraphics[width=0.98\columnwidth]{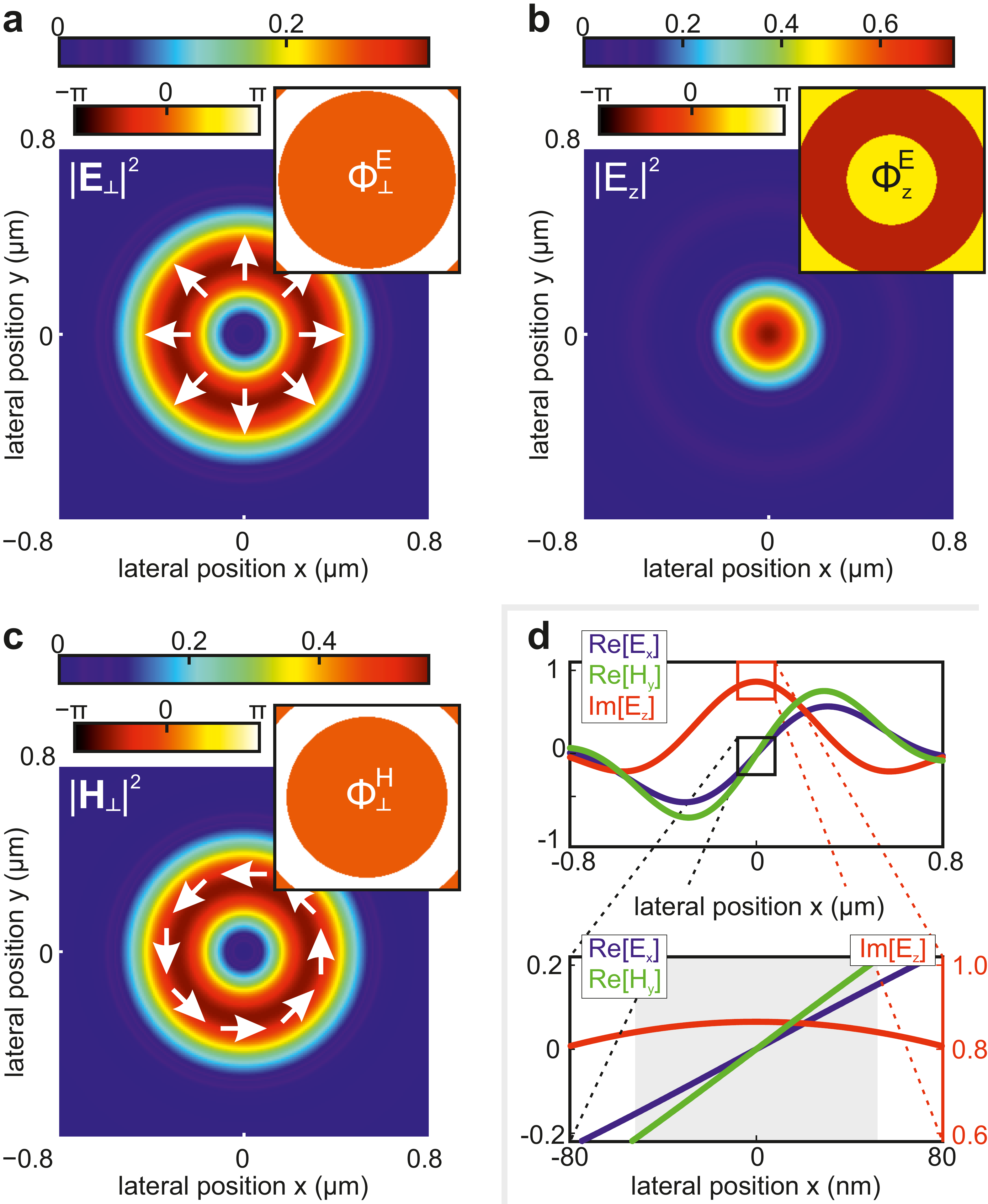}
\caption{Theoretical field intensity distributions and relative phases of a tightly focused radially polarized beam ($\lambda=652\text{ nm}$). Experimental parameters are taken into account (see Methods section). \textbf{a} The transverse (radial) electric field intensity $\left|\textbf{E}_{\bot}\right|^{2}=\left|E_{x}\right|^{2}+\left|E_{y}\right|^{2}$, \textbf{b} the longitudinal electric field intensity $\left|\textbf{E}_{z}\right|^{2}$, and \textbf{c} the transverse (azimuthal) magnetic field $\left|\textbf{H}_{\bot}\right|^{2}=\left|H_{x}\right|^{2}+\left|H_{y}\right|^{2}$ are all normalized to the maximum value of the total field intensity, $I_{tot}=\left|\textbf{E}\right|^2$ + $\left|\textbf{H}\right|^2$ (Gaussian units). \textbf{d} Cross-sections of the focal fields along the $x$-axis. Close to the center (gray area in the lower image), the transverse field amplitudes Re[$E_{x}$] and Re[$H_{y}$] are linearly dependent on the position, while the longitudinal field Im[$E_{z}$] is approximately constant.}
\label{fig_focal_plane}
\end{figure}

\begin{figure}
\includegraphics[width=0.98\columnwidth]{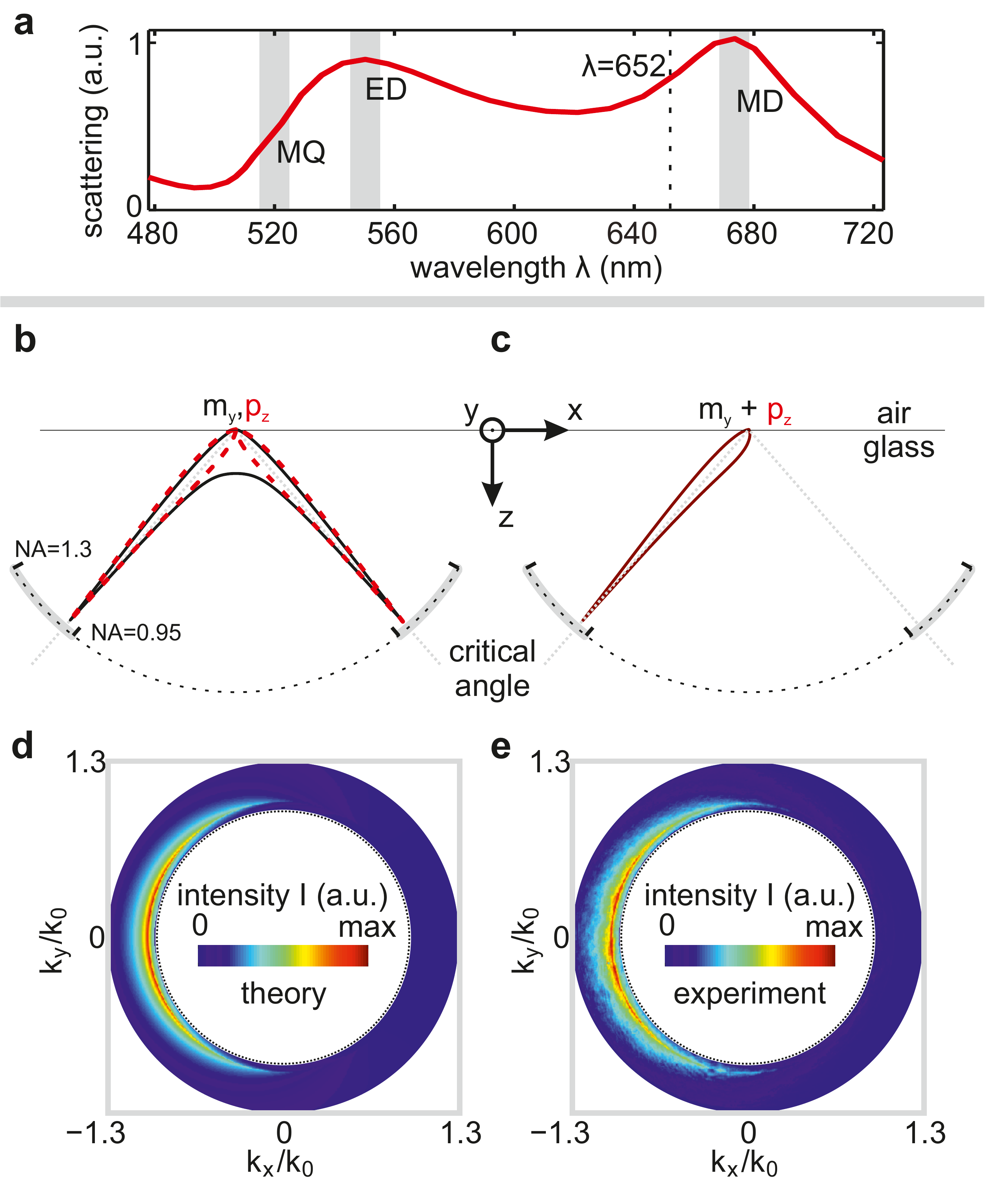}
\caption{Scattering of a silicon nanoantenna sitting on a dielectric interface. \textbf{a} Simulated scattering cross-section (linearly polarized Gaussian beam used for excitation) of a silicon sphere with radius $r=92\text{ nm}$; only the forward scattering efficiency into the angular region within NA~$\in[0.95,1.3]$ is considered to match the experimental detection scheme (see gray arcs in \textbf{b}, \textbf{c} and far-field patterns in \textbf{d}, \textbf{e}). In the visible range, the nanosphere supports magnetic dipole ($\lambda_{\text{MD}}\approx670\text{ nm}$), electric dipole ($\lambda_{\text{ED}}\approx550\text{ nm}$), and magnetic quadrupole ($\lambda_{\text{MQ}}\approx520\text{ nm}$) resonances. At the excitation wavelength of $\lambda=652\text{ nm}$ the magnetic and electric dipole moments are $\pi/2$ out-of-phase with respect to each other. \textbf{b} Emission of a longitudinal electric dipole $p_{\text{z}}$ (see dashed red line) and a transverse magnetic dipole $m_{\text{y}}$ (see black line) into the glass substrate. \textbf{c} In-phase far-field interference of $p_{\text{z}}$ and $m_{\text{y}}$ results in strong directivity. Comparison of \textbf{d} a calculated far-field pattern (interference of $p_{\text{z}}$ and $m_{\text{y}}$) and \textbf{e} a measured back-focal-plane image, retrieved at an antenna position on the $x$-axis $140 \text{ nm}$ away from the center of the beam.}
\label{fig_spectrum}
\end{figure}

\begin{figure*}[htbp]
\centerline{\includegraphics[width=1.66\columnwidth]{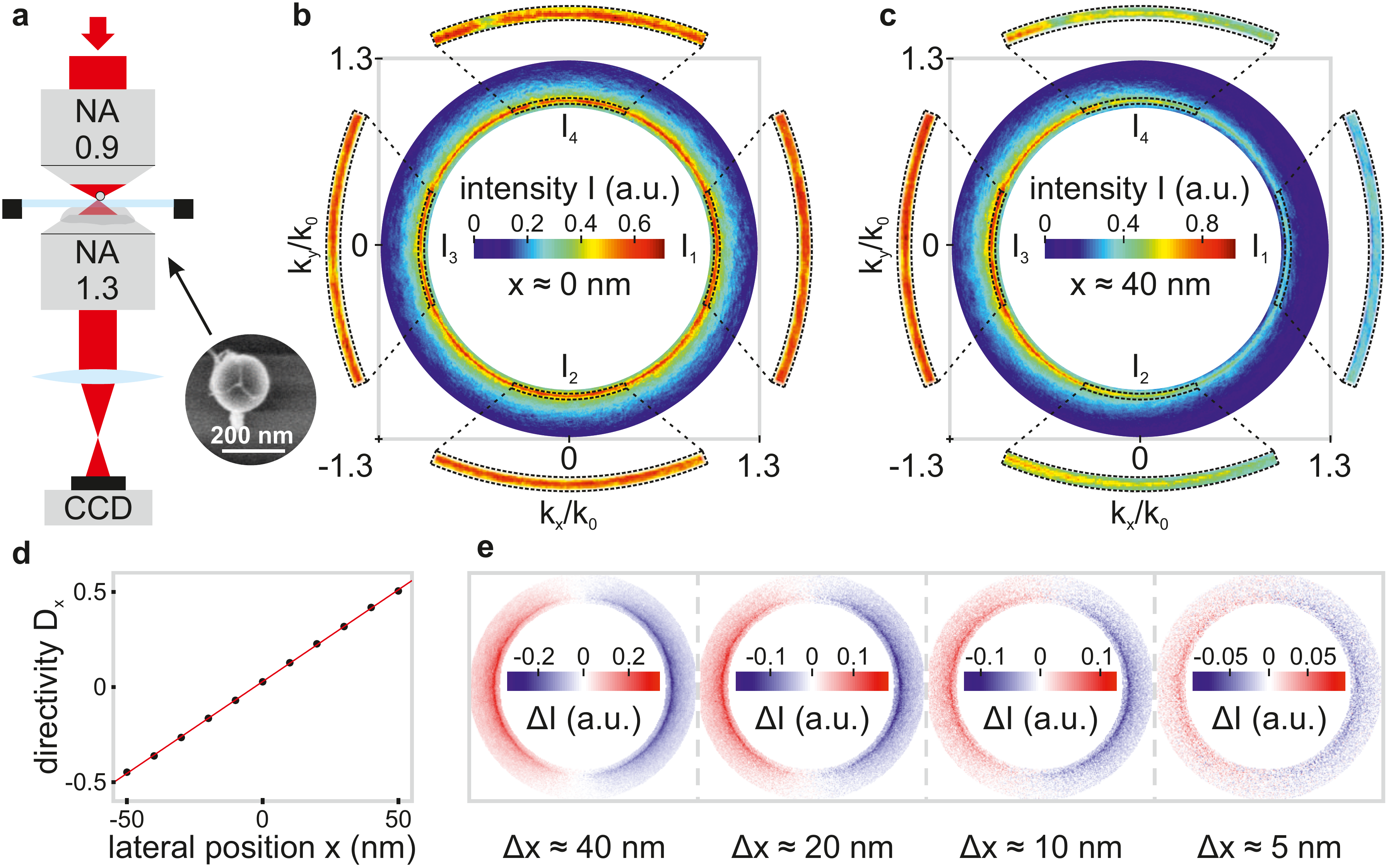}}
\caption{\label{fig_setup}Setup and experimental results. \textbf{a} A radially polarized collimated beam is tightly focused by a microscope objective of NA 0.9 onto a sub-wavelength silicon antenna sitting on a glass substrate (see inset). The light emitted into the angular regime with NA~$\in[0.95,1.3]$ is collected by an oil-immersion type microscope objective (NA~$=1.3$). The back-focal plane is imaged onto a CCD camera. \textbf{b} Acquired far-field intensity distribution $I(k_{x},k_{y})$ for the antenna placed on the optical axis and \textbf{c} for the antenna off-axis, displaced laterally by $ x\approx40\text{ nm}$. The dashed black lines and the magnified insets indicate four regions with averaged intensities $I_{1}$, $I_{2}$, $I_{3}$ and $I_{4}$. \textbf{d} Directivity parameter $D_{x}$ versus the antenna's position along the $x$-axis; the slope of the curve ($0.01/\text{nm}$) defines the sensitivity of the measurement to the antenna displacement. \textbf{e} Far-field intensity difference images for four antenna positions along the $x$-axis (left to right: $\Delta x \approx 40\text{ nm}$, $\Delta x \approx 20\text{ nm}$, $\Delta x \approx 10\text{ nm}$, and $\Delta x \approx 5\text{ nm}$ away from the optical axis). The left image corresponds to the intensity difference $\Delta I(k_{x},k_{y})$ between \textbf{b} and \textbf{c}.}
\end{figure*}

\textbf{Excitation scheme}. It was shown that the simultaneous excitation of transverse electric and magnetic resonances of a high-refractive-index dielectric nanoparticle may yield enhanced or suppressed forward/backward scattering due to their interference \cite{Kerker1983,Garcia-Camara2011,Fu2013,Alaee2015}. However, by carefully structuring the excitation field three-dimensionally and thus exciting also longitudinal particle modes \cite{Wozniak2015}, the scattering pattern can be tailored to achieve lateral directivity in the far-field. For example, a tightly focused radially polarized beam features a promising three-dimensional focal field with cylindrical symmetry \cite{Quabis2000,Youngworth2000}. Figures~\ref{fig_focal_plane}a-c show its electric and magnetic field intensity distributions and the corresponding phases, calculated by vectorial diffraction theory while taking into account the experimental parameters \cite{Richards1959,L.NovotnyandB.Hecht2006}. 
Apart from the transverse (in-plane) radially polarized electric field $\textbf{E}_{\bot}=(E_{x},E_{y})$, a strong longitudinal component $E_{z}$ is formed, reaching its maximum amplitude on the optical axis. In contrast, the magnetic field $\textbf{H}_{\bot}=(H_{x},H_{y})$ is purely transverse and azimuthally polarized. In close vicinity to the optical axis, $E_{z}$ exhibits a phase delay of $\Delta\phi_{z}=\pm\pi/2$ with respect to the transverse field components, and the electric and magnetic fields can be approximated by
\begin{subequations}\label{eqn_fields}
\begin{align}
\textbf{E}(x,y)&\propto xE_{\bot}^{0}\hat{e}_{x}+yE_{\bot}^{0}\hat{e}_{y}+iE_{z}^{0}\hat{e}_{z} \text{,}\\
\textbf{H}(x,y)&\propto -yH_{\bot}^{0}\hat{e}_{x}+xH_{\bot}^{0}\hat{e}_{y} \text{.}
\end{align}
\end{subequations}
Here, $E_{\bot}^{0}$, $E_{z}^{0}$, $H_{\bot}^{0}$ are real valued amplitudes of the transverse electric, longitudinal electric and transverse magnetic field components respectively, and $(x,y)$ are Cartesian coordinates in the focal plane. Without loss of generality, the point in time is chosen such that the transverse field components $E_{x}$, $E_{y}$, $H_{x}$ and $H_{y}$ are real, and the longitudinal component $E_{z}$ is imaginary, due to the aforementioned phase delay of $\pi/2$. For the chosen beam parameters (see Methods section), we estimate Eq.~\ref{eqn_fields} to be valid within the region up to $50\text{ nm}$ away from the optical axis (see gray area in Fig.~\ref{fig_focal_plane}d). In this limited range, the transverse electric and magnetic fields are linearly dependent on the coordinates $x$ and $y$, while $E_{z}$ is assumed to be approximately constant. In order to adopt this linear position dependence of the transverse electromagnetic field for position sensing, a sub-wavelength antenna, capable of incorporating the local field in its far-field emission pattern, is required to localize the antenna unambiguously by its (directional) scattering pattern recorded in the far-field. In the following, we discuss a silicon nanosphere (radius $r=92\text{ nm}$), whose spectrum in the visible range was experimentally investigated previously \cite{Wozniak2015}, and we explain how its far-field emission pattern is governed by its position.\\ \\
\textbf{Tailored directional scattering}. Figure \ref{fig_spectrum}a shows the scattering cross-section of the antenna sitting on a glass substrate, simulated using the finite-difference time-domain method (similar to ref.~\cite{Wozniak2015}). 
Here, only the forward scattering efficiency into the angular region within NA~$\in[0.95,1.3]$ is considered to match the experimental detection scheme described below (see also Fig.~\ref{fig_spectrum}b-e). In the visible spectrum, the silicon antenna supports three pronounced resonances, the magnetic dipole ($\lambda_{\text{MD}}\approx670\text{ nm}$), the electric dipole ($\lambda_{\text{ED}}\approx540\text{ nm}$) and the magnetic quadrupole ($\lambda_{\text{MQ}}\approx515\text{ nm}$) \cite{Wozniak2015}. For wavelengths above \mbox{$600\text{ nm}$}, the weak contribution of the magnetic quadrupole can be completely neglected \cite{Wozniak2015}, and the antenna can be approximated by a point-like dipole (electric and magnetic). Assuming that the dipole moments are proportional to the respective local field vectors, $\textbf{p}\propto\textbf{E}$ and $\textbf{m}\propto\textbf{H}$, we yield the position-dependent dipole moments $\textbf{p}\propto xE_{\bot}^{0}\hat{\textbf{x}}+yE_{\bot}^{0}\hat{\textbf{y}}+iE_{z}^{0}\hat{\textbf{z}}$ and $\textbf{m}\propto -yH_{\bot}^{0}\hat{\textbf{x}}+xH_{\bot}^{0}\hat{\textbf{y}}$. The aim of our experimental concept is to achieve highly position-sensitive far-field directivity caused by the interference of the, in first approximation, constant $z$-oriented electric dipole $p_{z}$ and the position-dependent transverse components of the magnetic dipole \mbox{$m_{x}$ and $m_{y}$}. The influence of the transverse electric dipole components $p_{x}$ and $p_{y}$ will be proven to be negligible later on.\\
In Fig.~\ref{fig_spectrum}b, the far-field intensities of a $z$-oriented electric dipole (see dashed red line) and a $y$-oriented magnetic dipole (see black line) emitted into the glass substrate are depicted. Here, we consider the electric and magnetic dipole moments to exhibit the same strength. If the dipole moments are in phase, the interference of both far-fields yields a remarkably strong lateral directivity (see Fig.~\ref{fig_spectrum}c). Figure \ref{fig_spectrum}d shows the corresponding calculated $k$-spectrum in the experimentally accessible region within NA~$\in[0.95,1.3]$. At this point, the relative phase between the longitudinal and the transverse field components ($\Delta\phi_{z}=\pm\pi/2$, see Fig.~\ref{fig_focal_plane}) of the excitation beam needs to be considered. If the electric and magnetic dipoles oscillate $\pi/2$ out-of-phase, no directivity would be observed in the far-field  because their symmetric far-field intensity distributions add up. Hence, an additional phase of $\pi/2$ is required to compensate for $\Delta\phi_{z}$. Since the relative phase between a dipole moment and its respective excitation field ($\Delta\phi_{MD}$ for the magnetic, $\Delta\phi_{ED}$ for the electric field) depends on the wavelength, we can compensate for $\Delta\phi_{z}$ by carefully choosing the wavelength of the incoming light with respect to the spectral positions of the electric and magnetic dipole resonances. From simulation, we retrieve the relative phase between the electric and magnetic dipole moment to be $\Delta \phi=\Delta\phi_{MD}-\Delta\phi_{ED}=\pi/2$ for a wavelength of $\lambda=652\text{ nm}$ (see the Supplementary Material). Using this wavelength for excitation, we expect to achieve strongly directional scattering at antenna positions where the longitudinal electric and transverse magnetic fields overlap. For experimental verification, a measured far-field image is plotted in Fig.~\ref{fig_spectrum}e. The image was retrieved by placing the antenna on the $x$-axis approximately $140\text{ nm}$ away from the center of the beam in the focal plane, effectively obtaining longitudinal electric and transverse magnetic dipole moments of equal strength. The strong directivity proves that the compensation of the relative phase has been successful, and the very good overlap of $94\%$ with the theoretical pattern suggests that in first-order approximation the transverse electric dipole moments, not taken into account here, can indeed be neglected (details can be found in the Supplementary Material).\\ 
In short, we optimized, the polarization distribution and the wavelength of our excitation beam to achieve strongly directional emission depending on the position of a single silicon nanoantenna relative to the beam's optical axis. The underlying principle causing the directivity is the simultaneous and in-phase excitation of a longitudinal electric and a transverse magnetic dipole moment.\\
\textbf{Experimental implementation and calibration}. A sketch of the experimental setup is depicted in Fig.~\ref{fig_setup}a (for more details see \cite{Banzer2010a}).
The collimated incoming radially polarized beam ($\lambda=652\text{ nm}$) was tightly focused by a microscope objective with a numerical aperture (NA) of 0.9 onto the silicon nanosphere sitting on a glass substrate (see electron micrograph in Fig.~\ref{fig_setup}a), which was positioned precisely within the focal plane by a 3D piezo-stage. A second microscope objective (oil-immersion type, NA~$=1.3$) below the substrate collected both, the transmitted beam and the forward scattered light. Imaging the back-focal plane of the second microscope objective onto a CCD camera enabled acquisition of the intensity distribution emitted into the far-field and grants access to the angular spectrum of the scattered field (see examples in Fig.~\ref{fig_spectrum}e and Fig.~\ref{fig_setup}b-c). Similar to \cite{Neugebauer2014}, only the region of NA~$\in[0.95,1.3]$ was considered, where the scattered light can be detected without interfering with the transmitted beam.\\
To retrieve the position of the antenna from the back-focal plane images, we average the measured intensity over four small regions in $k$-space (black dotted lines in Figs.~\ref{fig_setup}b-c) \cite{Neugebauer2015}, resulting in four averaged intensity values $I_{1}$, $I_{2}$, $I_{3}$, and $I_{4}$. The size and position of these four regions was chosen to include only the strongest change of the far-field intensity for a antenna shift along the $x$- or $y$-axis. The normalized intensity-differences $D_{x}=(I_{3}-I_{1})/\tilde{I}$ and $D_{y}=(I_{2}-I_{4})/\tilde{I}$, with $\tilde{I} = (I_{1}+I_{2}+I_{3}+I_{4})/2$, represent directivity parameters, which are linear functions of the antenna position (see Supplementary Material). \\
In order to compensate for experimental imperfections such as beam aberrations or deviations from the ideal antenna shape, the measurement approach requires initial calibration, for which we placed the antenna centrally in the focus. At this position, the far-field distribution of the scattered light is expected to be cylindrically symmetric, since only a longitudinal electric dipole moment can be excited (see Fig.~\ref{fig_setup}b) \cite{Neugebauer2014,Wozniak2015}. From this reference point, the antenna was scanned across the focal plane ($100\text{ nm}\times 100\text{ nm}$), with a step-size of $10\text{ nm}$. For each position, an image of the back-focal plane was acquired and the corresponding values of $D_{x}$ and $D_{y}$ were determined. The whole procedure was repeated 40 times and the measured directivity parameters were averaged, in order to decrease the influence of the instability of our setup (position uncertainty of $\pm5\text{ nm}$). Thereupon, linear equations were fitted to the averaged directivity parameters $D_{x} (x,y)$ and $D_{y}(x,y)$ (see Eq.~\ref{eqn_calib} in the Methods section), which allow for retrieving the antenna position from individual back-focal plane images. As an example, Fig.~\ref{fig_setup}d shows the averaged directivity parameter $D_{x}$ plotted against the $x$-coordinate and the corresponding linear fit.\\ \\ \\
\textbf{Lateral Resolution}. In order to demonstrate the accuracy in the measurement of the antenna position, which can be achieved with a single camera shot, far-field images for different antenna positions are analyzed. To this end, we normalize the intensity maps recorded in each back-focal plane to the intensity $\tilde{I}$ and calculate the difference to a reference image, which corresponds to the antenna sitting on the optical axis (see Fig.~\ref{fig_setup}b). For the demonstration of this inherently two-dimensional localization technique, we show results for $x$-displacements only. In Fig.~\ref{fig_setup}e we depict four post-selected difference-images for the antenna being placed on the $x$-axis, for which our calibration measurement indicated relative positions of $\Delta{x} \approx 40\text{ nm}$, $20\text{ nm}$, $10\text{ nm}$, and $5\text{ nm}$ ($\Delta y\approx 0\text{ nm}$). For a relatively large displacement of $\Delta x\approx 40\text{ nm}$, the difference-image corresponding to the difference between Fig.~\ref{fig_setup}b and Fig.~\ref{fig_setup}c yields a very good signal-to-noise ratio (SNR). Even for a small displacement of only $\Delta x\approx 5\text{ nm}$, the difference-image reveals predominately negative values on the left side ($k_{x}<0$) and positive values on the right side ($k_{x}>0$). However, the SNR decreases with shorter distances $\Delta x$. The theoretical limit of our resolution is determined by the derivatives (slopes) of $D_{x}(x,y)$ and $D_{y}(x,y)$, the intensity noise of an individual camera pixel, and the actual number of pixels in each integration region. Our calculations yield that a position uncertainty below 2 \AA{}ngstrom could be achieved. More details and an actual experimental example can be found in the Supplementary Material. However, a direct proof of this accuracy would require a highly stabilized setup including a piezo-stage with \AA{}ngstrom precision.
\section{Discussion}
In summary, we experimentally demonstrated that the simultaneous and phase-adapted excitation of longitudinal electric and transverse magnetic dipole modes of a high-refractive-index nanosphere yields extraordinarily strong directionality. Especially the spectral tuning of the relative phase between both dipole modes in combination with the appropriate choice of a three-dimensional focal field pattern enabled highly position-sensitive transverse scattering directionality. We utilized the approach as a novel technique for single-shot lateral position sensing, achieving localization accuracies down to a few \AA{}ngstrom, which is comparable to other state-of-the-art localization methods presented in literature \cite{Rohrbach2002,Nugent-Glandorf2004,Bon2015}. Our technique could be applied for the stabilization of samples, for instance in super-resolution microscopy. Furthermore, since the directionality is also present in the super-critical regime (NA~$>1$), evanescent coupling to waveguide modes will allow for on-chip detection of the directional scattering and, hence, of the lateral position of the sample. Finally, future studies might demonstrate, that antenna design and size as well as the excitation field can be optimized to achieve an even stronger dependence of the directionality on the particle position, which would allow for sub-\AA{}ngstrom localization accuracies. 

\section{Methods}
\begin {footnotesize}  \textbf{Experimental setup.} A tunable light source (NKT Photonics SuperK Extreme \& SpectraK Dual) emits a linearly polarized Gaussian beam at a wavelength of $652\text{ nm}$, which is converted into a radially polarized beam by a liquid-crystal polarization converter (q-plate) \cite{Marrucci2006,Slussarenko2011}. The beam with radius $w_{0}=1.26\text{ mm}$ is then guided into a microscope objective with NA~$=0.9$ and an entrance aperture radius of $1.8\text{ mm}$ (Leica HCX PL FL 100x/0.90 POL 0/D). A single spherical silicon nanoparticle with radius $r=92\text{ nm}$ on a glass substrate is scanned through the focal plane by a high-precision three-dimensional piezo-stage (PI P-527) and the transmitted light is collected with an oil-immersion objective with $\text{NA}=1.3$ (Leica HCX PL FLUOTAR ×100/1.30 OIL). The angular intensity distribution of the transmitted light is detected by imaging the back-focal plane of the oil-immersion objective onto a CCD camera (The Imaging Source DMK 23U618). The four solid angles corresponding to $I_{1}$, $I_{2}$, $I_{3}$ and $I_{4}$ (see dashed lines in Figs.~\ref{fig_setup}b-c and d) are defined by an azimuthal angular range of $\Delta\Phi=45^\circ$ and by NA~$\in[0.98,1.02]$. \\ \\
\textbf{Calculation of the far-field distribution.} We make use of the cylindrical symmetry of the beam and, without loss of generality, only consider antenna positions along the $x$-axis. Therefore, only the longitudinal electric ($p_{z}$) and transverse magnetic ($m_{y}$) dipole moments need to be considered. The transverse electric (s-polarized) and transverse magnetic (p-polarized) far-field distributions ($r\gg \lambda$) emitted into the dielectric substrate (refractive index $n=1.5$) are expressed as \cite{L.NovotnyandB.Hecht2006}
\begin{align}\label{eqn_far1}
E_{p}^{ED,z}&=-Ct_{p}\frac{k_{\bot}}{k_{0}}p_{z}\text{,}\\\label{eqn_far2}
E_{s}^{ED,z}&=0\text{,}\\ \label{eqn_far5}
E_{p}^{MD,y}&=\frac{C}{c_{0}}t_{p}\frac{k_{x}}{k_{\bot}}\,m_{y}\text{,}\\\label{eqn_far6}
E_{s}^{MD,y}&=-\frac{C}{c_{0}}t_{s}\frac{\sqrt{k_{0}^2-k_{\bot}^2}k_{y}}{k_{0}k_{\bot}}\,m_{y}\text{,}
\end{align}
with
\begin{align}\nonumber
C= e^{i k_{0} n r} \frac{k_{0}^{2}\sqrt{k_{0}^{2} n^{2}-k_{\bot}^{2}}}{4 \pi r \epsilon _{0} \sqrt{k_{0}^{2}-k_{\bot}^{2}}} e^{i \sqrt{k_{0}^{2} -k_{\bot}^{2}}d}\text{,}
\end{align}
the Fresnel coefficients for transmission $t_{p}$ and $t_{s}$, the wave-number in vacuum $k_{0}=2\pi/\lambda$, the transverse component of the $k$-vector $k_{\bot}=(k_{x}^2+k_{y}^{2})^{1/2}$ and the vacuum speed of light $c_{0}$. Comparison between theoretical and experimental far-field patterns enables estimating the distance between the effective point-like dipole and the interface, $d=70\text{ nm}$. The emission patterns in Fig.~\ref{fig_spectrum}b-c are calculated using Eq.~\ref{eqn_far1}-\ref{eqn_far6}, whereby we considered a similar strength for both dipole moments $p_{z}=m_{y}/c_{0}$ to achieve maximum directivity. Taking into account the aplanatic microscope objective, an additional energy conservation factor proportional to $(k_{0}^2n^2-k_{\bot}^2)^{-1/2}$ is introduced for Fig.~\ref{fig_spectrum}d \cite{L.NovotnyandB.Hecht2006}. \\ \\
\textbf{Calibration measurement.} The directivity parameters $D_{x}$ and $D_{y}$ are linear functions of the lateral antenna position $x$ and $y$ respectively. Thus, we fit a system of two linear equations to the averaged calibration measurement data, resulting in
\begin{align}\label{eqn_calib}
\begin{pmatrix} x\\y
\end{pmatrix}=
\begin{pmatrix}
103.4 \text{nm} & 4.8 \text{nm}  \\
-2.5 \text{nm}  & 94.8 \text{nm}
\end{pmatrix}
\begin{pmatrix} D_{x}\\ D_{y}
\end{pmatrix}+
\begin{pmatrix} -3.5  \text{nm} \\-12.7 \text{nm}
\end{pmatrix}
 \text{.}
\end{align}
Ideally, the matrix has non-zero values on its diagonal only. The small off-diagonal elements indicate a minor rotation of the coordinate system and, in addition, not entirely orthogonal directivity parameters $D_{x}$ and $D_{y}$. The rotation of the coordinates might stem from a misalignment of our camera with respect to the coordinate frame of the piezo-stage, while the non-orthogonal basis can be related to aberrations of the beam and asymmetries of the antenna (see electron micrograph in Fig.~\ref{fig_setup}a). The derivatives (slopes) of $D_{x}$ and $D_{y}$ define the sensitivity of the directivity to a displacement of the antenna. At the rim of the region of linearity, $50\text{ nm}$ away from the center, we already achieve a directivity $D_{x}=48\%$ ($D_{y}=52\%$) if the antenna is shifted in $x$-direction ($y$-direction).\\ \\
\textbf{Post-selection of difference images.}
The instability of our experimental setup causes an uncertainty of $\pm5\text{ nm}$ regarding the position of the particle relative to the beam. For this reason, we took $40$ individual images for each position set by the piezo-stage ($\Delta{x} \approx 40\text{ nm}$, $20\text{ nm}$, $10\text{ nm}$, and $5\text{ nm}$), and then post-selected the far-field images of which the directivity parameters $D_{x}$ and $D_{y}$ best represented the position set by the piezo-stage according to the calibration measurement (see Eq.~\ref{eqn_calib}). Finally, we calculated the difference images depicted in Fig.~\ref{fig_setup}e.
\end {footnotesize}\\ \\
\begin{acknowledgments}
We thank A. Rubano and L. Marrucci for the fabrication of the liquid crystal polarization converter (q-plate) utilized for the generation of the radially polarized beam, and T. Bauer and U. Mick for inspiring discussions. P. Banzer acknowledges financial support by the Alexander von Humboldt Foundation and the Canada Excellence Research Chair (CERC) in Quantum Nonlinear Optics.
\end{acknowledgments}
\bibliography{bib}
\newpage
\section{Supplemental Material}
\begin{figure}[h]
\centerline{\includegraphics[width=0.82\columnwidth]{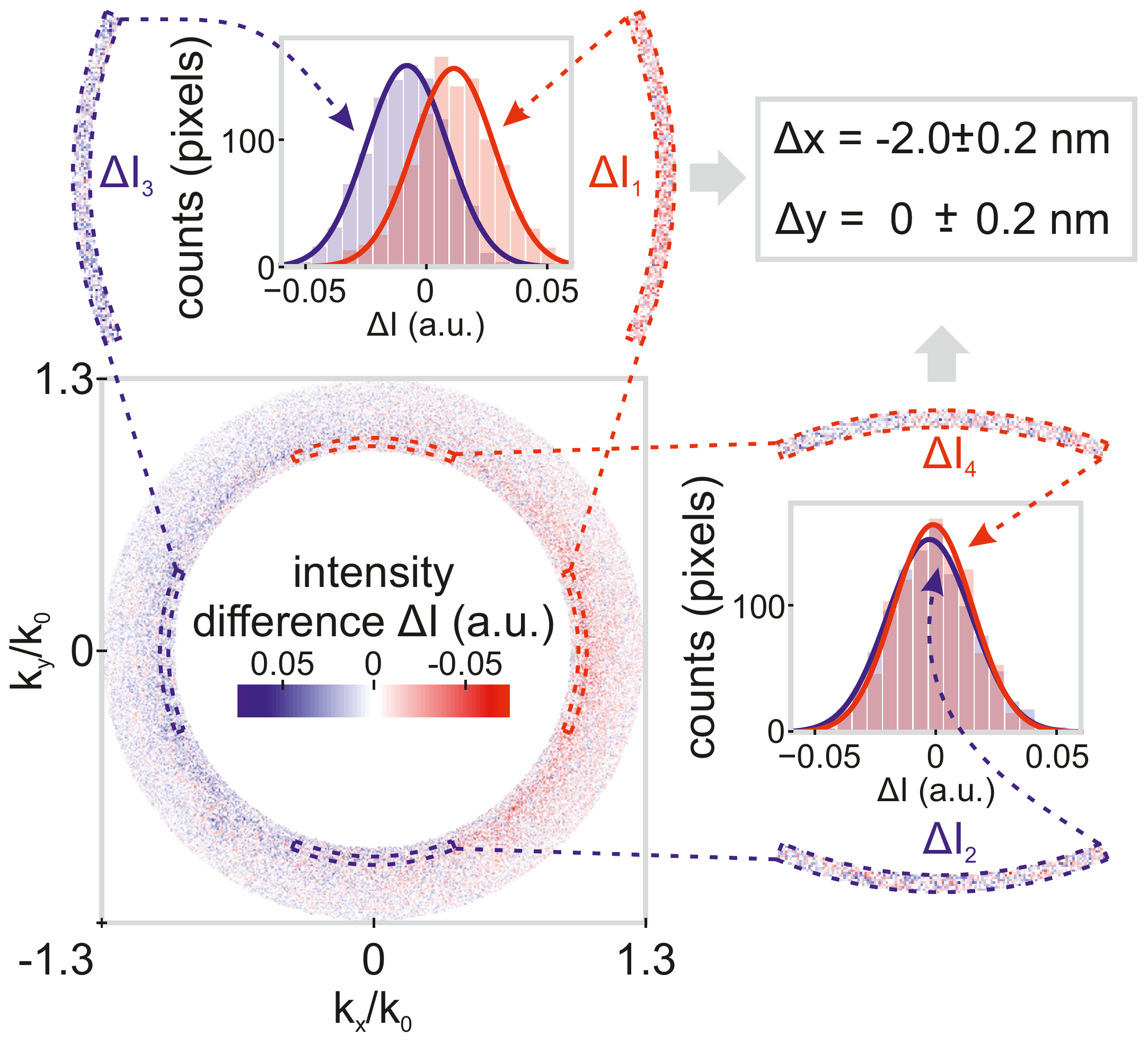}}
\caption{\label{fig_sub_resolution} Difference image of two back-focal plane images with similar intensity distributions ($\Delta x \approx -2\text{ nm}$, $\Delta y \approx 0\text{ nm}$). The statistics of the camera noise in each region (histograms shown as insets) can be used to determine the resolution of our experiment.}
\end{figure}

\textbf{Estimation of the resolution.}
In order to estimate the resolution of our position sensing experiment, we compare two post-selected nearly identical far-field images (see difference image in Fig.~\ref{fig_sub_resolution}) and calculate the average intensity differences for all four regions, $\Delta I_{1}=11\cdot10^{-3}$, $\Delta I_{2}=-10^{-3}$, $\Delta I_{3}=-8\cdot10^{-3}$, $\Delta I_{4}=-10^{-3}$. The corresponding standard deviations (\mbox{$\sigma_{i}\approx 25\cdot10^{-3}$ for $i\in[1,4]$}, see histograms plotted as insets in Fig.~\ref{fig_sub_resolution}) and the number of pixels in each region (1050 pixels), yield an uncertainty of the mean intensities of $\pm 10^{-3}$ for each intensity value. These results indicate that a relative shift of the antenna's position of $\Delta x = -2\pm 0.2 \text{ nm}$ and $\Delta y = 0 \pm 0.2 \text{ nm}$ was measured with an uncertainty in the \AA{}ngstrom regime. Hence, the two almost identical back focal plane images used for this estimation correspond to two antenna positions, which were different by only $2\pm 0.2\text{ nm}$.  
\\

\begin{figure}[htbp]
\centerline{\includegraphics[width=0.82\columnwidth]{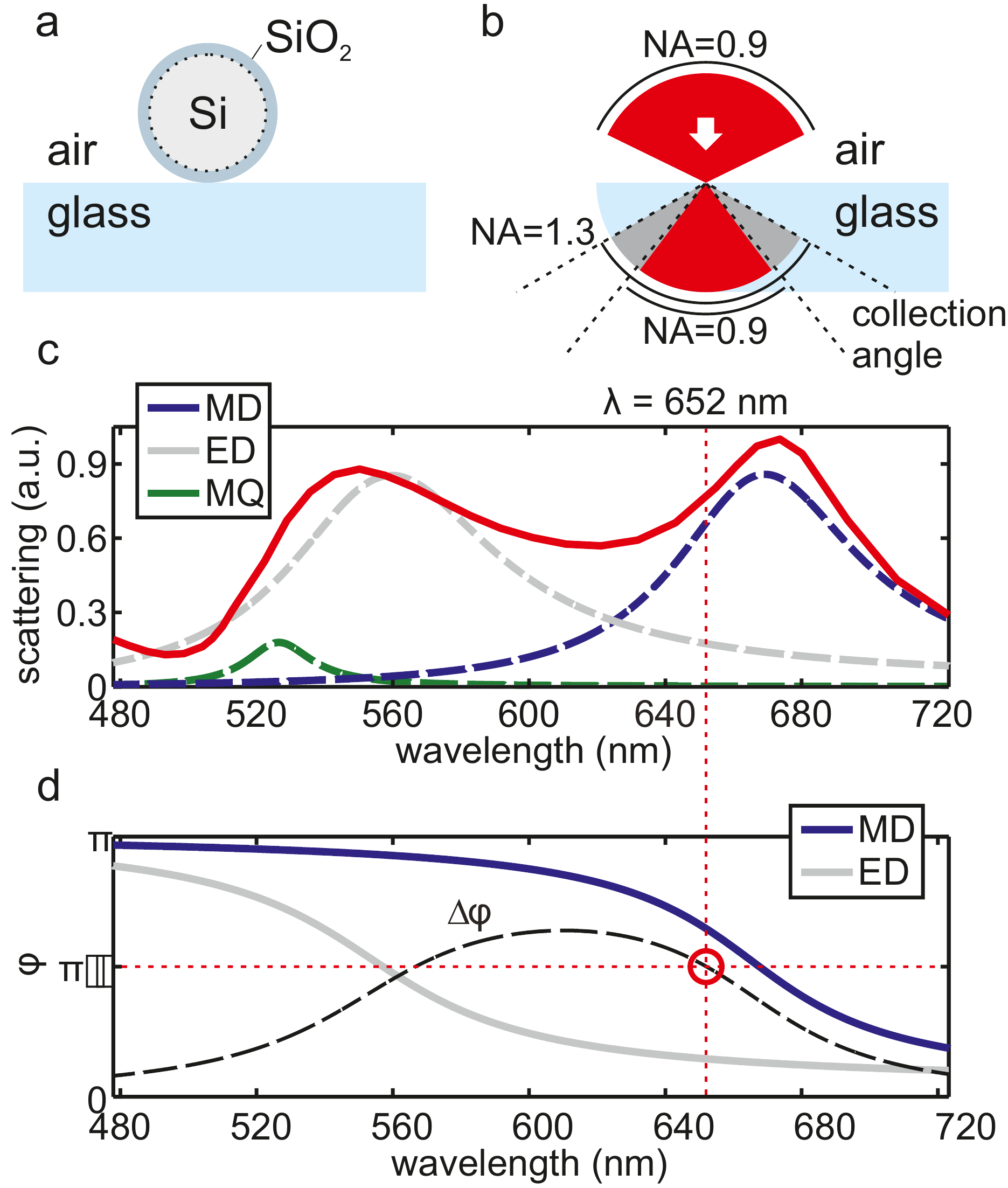}}
\caption{\label{fig_sub_spectrum} \textbf{a} Sketch of the Si antenna model with SiO$_{2}$ shell used for simulation. \textbf{b} Aperture angles of the microscope objectives used for focusing the incoming beam ($\text{NA}=0.9$) and collecting the scattered and transmitted light ($\text{NA}=1.3$). The actual collection angle is set to NA $\in [0.95,1.3]$. \textbf{c} Scattering cross-section as depicted in Fig.~2 of the main manuscript (see solid red line). A sum of three Lorenzians representing the magnetic dipole MD (dashed blue line), electric dipole ED (dashed gray line) and the magnetic quadrupole MQ (dashed green line) is fitted to the simulated spectrum. \textbf{d} Relative phases between the excitation field and the corresponding magnetic ($\phi_{MD}$, solid blue line) and electric ($\phi_{ED}$, solid gray line) dipole moments. At the chosen excitation wavelength of $652\text{ nm}$, the phase difference $\Delta \phi=\phi_{MD}-\phi_{ED}$ (dashed black line) is $\pi/2$.}
\end{figure}

\textbf{Phase retrieval from simulation.} Figure 2a of the main manuscript and Fig.~\ref{fig_sub_spectrum}c of this supplemental document show the scattering cross-section of the silicon (Si) antenna sitting on a glass substrate, simulated using the finite-difference time-domain method. Similar to refs.~\cite{Evlyukhin2012,Fu2013,Wozniak2015}, our Si nanosphere with an outer diameter of $185 \text{ nm}$ is modeled including a thin SiO$_{2}$ surface layer with an approximated thickness of $8\text{ nm}$ due to oxidation (see sketch in Fig.~\ref{fig_sub_spectrum}a).
\\
The material properties of Si and SiO$_{2}$ are adopted from the database of Palik \cite{Palik1995}. In general, Si has a high refractive index and a small extinction coefficient in the visible regime (e.g. $n_{Si} = 3.85-0.02i$ at $\lambda = 652 \text{nm}$). In order to adapt the simulation to the experiment, we consider a tightly focused linearly polarized Gaussian beam as a source (maximum NA $= 0.9$), and collect only the light scattered in the forward direction into the far-field with a polar collection angle set to NA $\in [0.95,1.3]$ (see sketch in Fig.~\ref{fig_sub_spectrum}b).\\
In the investigated spectral regime, a Si nanosphere of the chosen size placed on a substrate supports three pronounced resonances, the magnetic dipole ($\lambda_{\text{MD}}\approx660\text{ nm}$), the electric dipole ($\lambda_{\text{ED}}\approx540\text{ nm}$) and the magnetic quadrupole ($\lambda_{\text{MQ}}\approx515\text{ nm}$) \cite{Wozniak2015}. For this reason, the scattering spectrum (see red line in Fig.~\ref{fig_sub_spectrum}c) can be described, in first approximation, by the sum of three individual Lorentzian curves, fitted to the simulation (see Fig.~\ref{fig_sub_spectrum}c). The weak contribution of the magnetic quadrupole (dashed green line) can be neglected for wavelengths above \mbox{$600\text{ nm}$}. Therefore, only electric and magnetic dipole resonances need to be considered for the chosen excitation wavelength of \mbox{$652\text{ nm}$}. Now, we consider the relative phases of the magnetic and electric dipole moments with respect to their corresponding excitation fields. Following the Lorentz oscillator model \cite{Joe2006}, the relative phases of the electric ($\phi_{ED}$) and magnetic ($\phi_{MD}$) dipole moments depend on the excitation wavelength. Figure \ref{fig_sub_spectrum}d shows $\phi_{ED}$ (gray line), $\phi_{MD}$ (blue line), and the phase difference $\Delta \phi=\phi_{MD}-\phi_{ED}$ (see dashed black line). We find two wavelengths with $\Delta \phi = \pi/2$. However, we choose the wavelength $652\text{ nm}$, which is close to $\lambda_{\text{MD}}$ and guarantees a sufficient overlap between both types of dipoles (see dashed vertical red line). As mentioned above, the magnetic quadrupole can be neglected for the chosen wavelength. Another advantage of this choice of wavelength close to the magnetic resonance is the much higher efficiency, with which the magnetic dipole mode can be excited in comparison to its electric counterpart. This leads to comparable scattering signal strengths from both induced magnetic and electric dipole moments (even though the electric field is much stronger than the magnetic field close to the optical axis) and, therefore, stronger asymmetry upon interference. Consequently, an enhanced position dependence of the directionality is realized.\\

\textbf{Law of proportionality for the directivity parameters.}
The directivity parameters $D_{x}$ and $D_{y}$ correspond to differences of averaged intensity values recorded in the far-field. In order to derive equations describing the position dependence of $D_{x}$ and $D_{y}$, we first calculate the far-field intensity patterns $I(k_{x},k_{y})$ depending on the longitudinal electric dipole moment ($p_{z}\in \mathbb{R}$) and the transverse magnetic dipole moments ($m_{x}, m_{y}\in \mathbb{R}$). As mentioned in the main manuscript, we neglect any influence of the transverse electric dipole moments (see explanation below) as well as the magnetic quadrupole (see discussion above). Because of the cylindrical symmetry of the excitation beam and without loss of generality, we only consider antenna positions along the $x$-axis ($m_{x}=0$). The far-field intensity distribution of the light emitted into the glass substrate can be written as $I(k_{x},k_{y})=\left|E_{p}\right|^{2} + \left|E_{s}\right|^{2}$, with 
\begin{align}\label{eqn1} 
E_{p}= E_{p}^{ED,z}+E_{p}^{MD,y}\text{,}\\\label{eqn2}
E_{s}= E_{s}^{ED,z}+E_{s}^{MD,y}\text{.}
\end{align}
From Eqs.~2-5 in the Methods section of the main manuscript it follows
\begin{widetext}
\begin{align}\label{eqn3} 
I(k_{x},k_{y})= |Ct_{p}|^{2} \left[\left(\frac{k_{\bot}}{k_{0}}p_{z}\right)^{2}+\left(\frac{k_{x}}{c_{0}k_{\bot}}m_{y}\right)^{2}- 2\frac{k_{x}}{c_{0}k_{0}}p_{z}m_{y}\right] + |Ct_{s}|^{2} \left|\frac{\sqrt{k_{0}^2-k_{\bot}^2}k_{y}}{c_{0}k_{0}k_{\bot}}\,m_{y}\right|^{2} \text{.}
\end{align}
\end{widetext}
Experimentally, we only consider the far-field intensity close to the critical angle $I_{c}$, which implies $k_{\bot}\approx k_{0}$ and simplifies Eq.~\ref{eqn3} to
\begin{align}\label{eqn4} 
I_{c}(k_{x},k_{y})= |Ct_{p}|^{2} \left[p_{z}^{2}+\left(\frac{k_{x}}{c_{0}k_{0}}m_{y}\right)^{2}- 2\frac{k_{x}}{c_{0}k_{0}}p_{z}m_{y}\right] \text{.}
\end{align}
The difference between two intensity values with respect to the $y$-axis yields
\begin{align}\label{eqn5} 
I_{c}(-k_{x},k_{y})-I_{c}(k_{x},k_{y})= 4|Ct_{p}|^{2} \frac{k_{x}}{c_{0}k_{0}}p_{z}m_{y} \text{.}
\end{align}
Since only the transverse magnetic dipole depends on the position as $m_{y}\propto xH_{\bot}^{0}$ (see manuscript), it directly follows that $D_{x}\propto I_{c}(-k_{x},k_{y}) - I_{c}(k_{x},k_{y}) \propto x$. The result can also be extended to the two-dimensional case with \mbox{$D_{y}\propto I_{c}(k_{x},-k_{y}) - I_{c}(k_{x},k_{y})  \propto y$}.\\
 
\begin{figure}[!ht]
\centerline{\includegraphics[width=0.9\columnwidth]{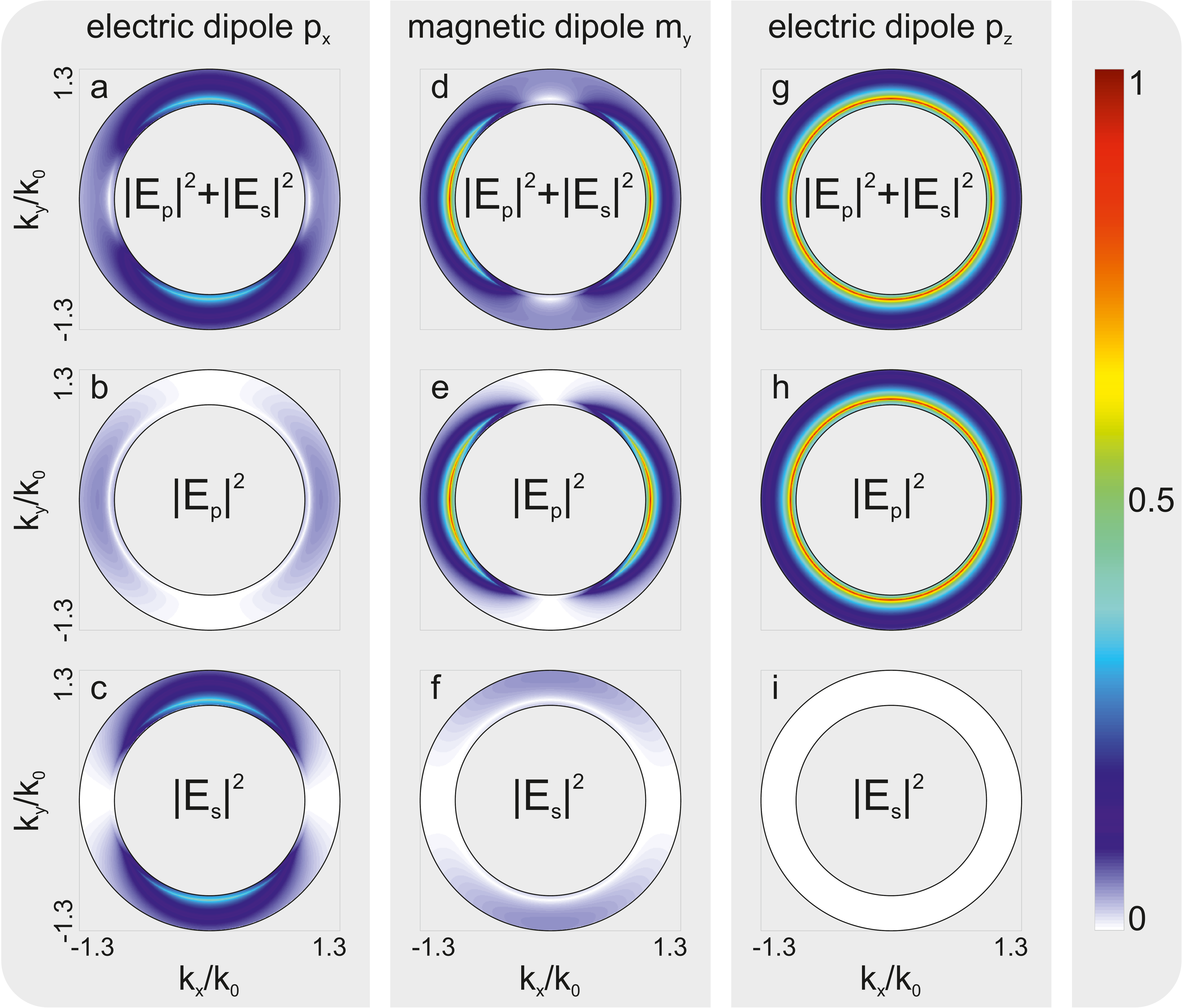}}
\caption{\label{fig_sub_farfield} Calculated far-field intensity patterns of the dipole moments $p_{x}$, $m_{y}$ and $p_{z}$ emitted into the glass substrate (distance between the effective dipole and the glass interface $d=70\text{ nm}$). Similar to the experiment, only the angular region within NA $\in\left[0.95,1.3\right]$ is considered. \textbf{a}, \textbf{d} and \textbf{g} illustrate the total far-field intensities $I=\left|E_{p}\right|^{2} + \left|E_{s}\right|^{2}$, with \textbf{b}-\textbf{c}, \textbf{e}-\textbf{f} and \textbf{h}-\textbf{i} depicting the individual far-field polarization components $\left|E_{p}\right|^{2}$ and $\left|E_{s}\right|^{2}$, respectively. All intensity distributions are normalized to the maximum of \textbf{g}.}
\end{figure}

\textbf{Negligibility of the transverse electric dipole.} As mentioned in the main manuscript, we expect the electric and magnetic dipole moments to be proportional to the respective local field vectors, $\textbf{p}\propto\textbf{E}$ and $\textbf{m}\propto\textbf{H}$. This includes the transverse electric dipole moments $p_{x}\propto E_{x}$ and $p_{y}\propto E_{y}$. However, for several reasons we can neglect the influence of the transverse electric dipole moments in first approximation.\\
First of all, even at the rim of the region of linearity, roughly $50 \text{ nm}$ away from the optical axis, the longitudinal electric field is still stronger than the transverse ones by a factor of 4 (see Fig.~1d in the manuscript). Second, with the given excitation wavelength close to the magnetic dipole resonance of the antenna, we expect to excite the transverse magnetic dipole moment with a higher efficiency than the transverse electric dipole moment (see Fig.~\ref{fig_sub_spectrum}). A third reason is based on the actual far-field emission patterns of the individual dipole moments plotted in Figs.~\ref{fig_sub_farfield}a-i.
Similar to ref.~\cite{Neugebauer2014}, the lateral directivity is linked to the transverse magnetic far-field component $E_{p}$. However, only $p_{z}$ is emitting exclusively as $E_{p}$ (see Eq.~3 of the main manuscript and Fig.~\ref{fig_sub_farfield}i of this supplemental document). The influence of the individual dipole moments on the directivity is therefore governed by the amount of light emitted into $E_{p}$. For each dipole we calculate the power of the transverse magnetic field by integrating over the distributions plotted in Fig.~\ref{fig_sub_farfield}b, e and h. The ratios between these power values for $p_{x}$, $m_{y}$ and $p_{z}$ equals to $1:6.4:14.5$. We conclude that the influence of transverse electric dipole moments can indeed be neglected in first approximation. In particular, its influence on the directivity parameters is very small.
\end{document}